\documentclass[%
superscriptaddress,
longbibliography,
amsmath,amssymb,
prf,
]{revtex4-1}
\usepackage[export]{adjustbox}
\usepackage{mathtools}
\usepackage{graphicx}% Include figure filessn
\usepackage{color}% Include figure files
\usepackage{xcolor}% Include figure files
\usepackage{wasysym}% Include figure files
\usepackage{dcolumn}% Align table columns on decimal point
\usepackage{bm}% bold math
\usepackage{hyperref}% add hypertext capabilities
\usepackage[mathscr]{euscript}
\graphicspath{{fig/}}
\definecolor{mygreen}{rgb}{0,0.5,0}
\definecolor{mybrown}{rgb}{0.65,0.16,0.16}

\def\epsd{\epsilon_\theta}
\def\dth{\delta \theta}

\def\beq {\begin{equation}}
\def\eeq {\end{equation}}
\def\beqa {\begin{eqnarray}}
\def\eeqa {\end{eqnarray}}
\def \bnum {\begin{enumerate}}
\def \enum {\end{enumerate}}
\def\bi {\begin{itemize}}
\def\ei {\end{itemize}}
\def \bdes {\begin{description}}
\def \edes {\end{description}}

\def\rel {R_{\lambda}}

\def\app{\approx}

\def\dho{\partial}
\def\la {\langle}
\def\ra {\rangle}

\def\diff{D_\phi}

\def\thetarms{\theta^\prime}

\def\scgx{|\dho\theta/\dho x|}

\def\scglarge{\thetarms/\eta}
\def\thetanr{\theta/\thetarms}
\begin{document}
%\setlength{\abovedisplayskip}{3pt}
%\setlength{\belowdisplayskip}{3pt}
%\preprint{APS/123-QED}
\title{Fractal iso-level sets in high-Reynolds-number scalar turbulence}
%%%%%%%%%%%%%%%%%%%%%%%%%%
\author{Kartik P. Iyer}
\affiliation{Tandon School of Engineering, New York University, New York, NY 11201, USA}
%%%%%%%%%%%%%%%%%%%%%%%%%%
\author{J\"{o}rg Schumacher}
\address{Institut f\"ur Thermo-und Fluiddynamik, Technische Universit\"at Ilmenau, Postfach 100565, D-98684 Ilmenau, Germany}
\affiliation{Tandon School of Engineering, New York University, New York, NY 11201, USA}
%%%%%%%%%%%%%%%%%%%%%%%%%%
\author{Katepalli R. Sreenivasan}
\email{krs3@nyu.edu}
\affiliation{Tandon School of Engineering, New York University, New York, NY 11201, USA}
\address{Department of Physics and the Courant Institute of Mathematical Sciences, New York University, New York, NY 10012, USA}
%%%%%%%%%%%%%%%%%%%%%%%%%%
\author{P. K. Yeung}
\address{Schools of Aerospace and Mechanical Engineering, Georgia Institute of Technology, Atlanta, GA 30332, USA}

\date{\today}% It is always \today, today,
\begin{abstract}
We study the fractal scaling of iso-levels sets of a passive scalar mixed by three-dimensional homogeneous and isotropic turbulence at high Reynolds numbers. The Schmidt number is unity. A fractal box-counting dimension $D_F$ can be obtained for iso-levels below about 3 standard deviations of the scalar fluctuation on either side of its mean value. The dimension varies systematically with the iso-level, with a maximum of about 8/3 for the iso-level at the mean; this maximum dimension also follows as an upper bound from the geometric measure theory. We interpret this result to mean that mixing in turbulence is always incomplete. A unique box-counting dimension for all iso-levels results when we consider the spatial support of the steep cliffs of the scalar conditioned on local strain; that unique dimension is about 4/3.
\end{abstract}
\maketitle
%\vspace{-5cm}
%\maketitle
\section{Introduction}
Consider a homogeneous and isotropic turbulence field in a periodic box at a high Reynolds number, generated by direct numerical simulations (DNS) of the Navier-Stokes (NS) equations. The turbulent field is maintained statistically stationary by supplying energy at a few low wave number shells. Into this turbulence field we introduce passive scalar fluctuations statistically uniformly, and allow them to evolve according to the advection diffusion equation along with the NS equations; the scalar field is maintained steady by means of a constant scalar gradient in one direction. For clarity, scalars are quantities that can be specified by their magnitude alone, and passive scalars do not influence the dynamics of turbulence that advects it. The diffusivity of the scalar is small and is equal to the viscosity of the fluid (i.e., the Schmidt number is unity). Modest amount of heat in air flows is a concrete example close enough to the situation we have in mind. The properties of passive scalar fields with a variety of Schmidt numbers have been explored in a few classical papers in the late 1940's to mid-50's \cite{obukhov1949,corrsin1951,batchelor1959}; a summary of the progress made since then, and references to important papers on the subject, can be found in \cite{sreenivasan1991PRS,warhaft2000,dimotakis2005,gotoh2013,sreeni2019}. 

Figure \ref{fig1.fig} shows a typical planar section of the passive scalar field just described. Its first conspicuous property is the presence of large scale fronts, often called ramp-cliff structure (see, e.g., Refs.~\cite{sreenivasan1991PRS,holzer}), ``cliffs" because of the tendency of the scalar to rise to the high concentration value rather abruptly while decaying to the lower concentration value rather gradually (``ramp"); across a cliff the nearly abrupt jump of the concentration of the scalar is on the order of magnitude of the entirety of the scalar difference available in the box. This latter is equal to the product of the mean scalar gradient, $G$, and the linear dimension of the box, $L_0$. These fronts occur even when the velocity field is turbulent and the scalar has the full-band of standard spectral shape that we have come to expect \cite{sreeni2019}. The existence of such sharp and large fronts endows the scalar field with certain types of anomaly studied most recently in \cite{iyer2018}. Briefly, we find that the scaling exponents of the scalar structure functions approach constant values even when the order of the structure function increases without bound. This behavior is unexpected from the classical point of view and is a property shared with model problems such as the Burgers equation for pressure-less velocity fields (for a review, see, Ref. \cite{falk}) and the Kraichnan model \cite{kraichnan1968} wherein the mixing velocity is a rapidly oscillating Gaussian field. 

The second property to which we draw attention is that such fronts consist of convolutions on many scales (see Fig.~\ref{fig1.fig}). For example, an enlarged view of Fig.~\ref{fig1.fig}(a) shows the same qualitative features of the front in Fig.~\ref{fig1.fig}(b), see the front indicated by AB. Indeed, the gradient of the passive scalar shows even more clearly that the front consists of many scales, and an enlargement of its part is similar to the entire scalar gradient field. This feature is displayed in Fig.~\ref{fig1.fig}(c) and Fig.~\ref{fig1.fig}(d). One can visually appreciate that the fronts contain convolutions over a number of scales. More specifically, any iso-level set for the concentration of the passive scalar contains contortions on many scales. Here, iso-level set means the set in three-dimensional space corresponding to fixed level (or threshold) of the scalar. An obvious expectation then is that a fractal-like description \cite{mandelbrot1977} holds for such iso-levels. It was first explored much more concretely in Ref.~\cite{sreenivasan1986,sreenivasan1991} and later by others, cited fully in a recent work \cite{dasi2007}. Our study here focusses on a detailed analysis of this property of scalar iso-level sets in relation to the ramp-cliff structure of the fronts. The study will be based on high-resolution DNS data of passive scalar turbulence, described next.

\section{Turbulence simulations}
We use data from pseudo-spectral DNS of homogeneous isotropic turbulence, computed on $4096^3$ grid points in a periodic cubical box of size $L_0=2\pi$. The passive scalar $(\Theta)$ is evolved in the same box using the advection diffusion equation in the presence of a uniform mean gradient ${\bm G} \equiv (G,0,0)$ along the $x$-direction, where $G \ne 0$ is a constant, such that $\Theta = \theta + Gx$ and $\theta$ is the scalar fluctuation field. The velocity field ${\bm u}$ is incompressible. The equations of motion are 
%---------------------------------------------------------------------------------
\begin{align}
{\bm \nabla}\cdot {\bm u}&=0\,,\label{mass}\\
\frac{\partial {\bm u}}{\partial t} + ({\bm u}\cdot{\bm \nabla}){\bm u} &= -{\bm \nabla}p + \nu {\bm\nabla}^2 {\bm u}+ {\bm f}\,,\label{mom}\\ 
\frac{\partial \theta}{\partial t} + ({\bm u}\cdot{\bm \nabla}) \theta &= D {\bm\nabla}^2 \theta -u_x G\,,\label{scalar}
\end{align}
%---------------------------------------------------------------------------------
with the large-scale forcing ${\bm f}$ sustaining a statistically stationary flow and $p$ the (kinematic) pressure field. The Schmidt number ${\rm Sc}=\nu/D=1$ and the Taylor microscale Reynolds number is $R_{\lambda}=650$. In total, we have used over $30$ essentially independent temporal snapshots spanning over $10$ eddy turnover times $T_E \equiv L/u^{\prime}$, where $u^\prime$ is the root-mean-square velocity fluctuation and $L$ is the integral scale with $L/L_0 \approx 0.2$. The ratio of the root-mean-square scalar fluctuation $\theta^\prime$ to the the maximum available mean scalar difference in the box is $\theta^\prime/GL_0 \approx 0.2$. The spectral resolution is chosen such that the ratio of grid spacing $\Delta$ to the Kolmogorov length $\eta$ is given by $\Delta/\eta=1.1$, where $\eta=(\nu^3/\langle\epsilon\rangle)^{1/4}$ and $\langle\epsilon\rangle$ is the mean kinetic energy dissipation rate. An inertial subrange in agreement with Kolmogorov's 4/5-ths law is established for scales between $30\eta$ and $300\eta$. In total, a linear scale range from $\eta$ up to about $2000 \eta$ is captured. For further details on the numerical resolution, inertial range properties and statistical convergence, see Refs.~\cite{yeung2012,iyer2018}. 

\section{Box-counting analysis of different scalar iso-level sets} \label{sec3}
Fractals are spatial objects that follow a self-similar scaling in the form of power laws \cite{mandelbrot1977,sreenivasan1991}. An experimental realization of a fractal requires a significant range of scales. In a homogenous turbulent flow, the available scale range varies as $L/\eta \approx {\rm Re}^{3/4}$ where the flow Reynolds number is given by ${\rm Re}=u^{\prime}L/\nu$. A fractal scaling with a box-counting dimension $D_F$ exists if the number $N(r)$ of boxes with edge-length $r$ cover an object, in this instance an iso-level set, with the scaling law
%---------------------------------------------------------------------------------
\begin{equation}
N(r) = N(L) \left(\frac{r}{L}\right)^{-D_F}\,
\end{equation}    
%---------------------------------------------------------------------------------
for some significant range of scales. The early experiments \cite{sreenivasan1986} were for inhomogeneous flows, typically at modest Reynolds numbers, with some attendant uncertainties of scaling. Here, we have on hand fully resolved three-dimensional data that span a range of scales that is three orders of magnitude larger. The scale range of the simulation data is also much larger than those of previous simulations such as Refs.~\cite{brandenburg92,sangil2001,JS05a}.

%%%%%%%%%%%%%%%%%%%%%%%%%%%5
\begin{figure*}
\includegraphics [width=0.98\textwidth]{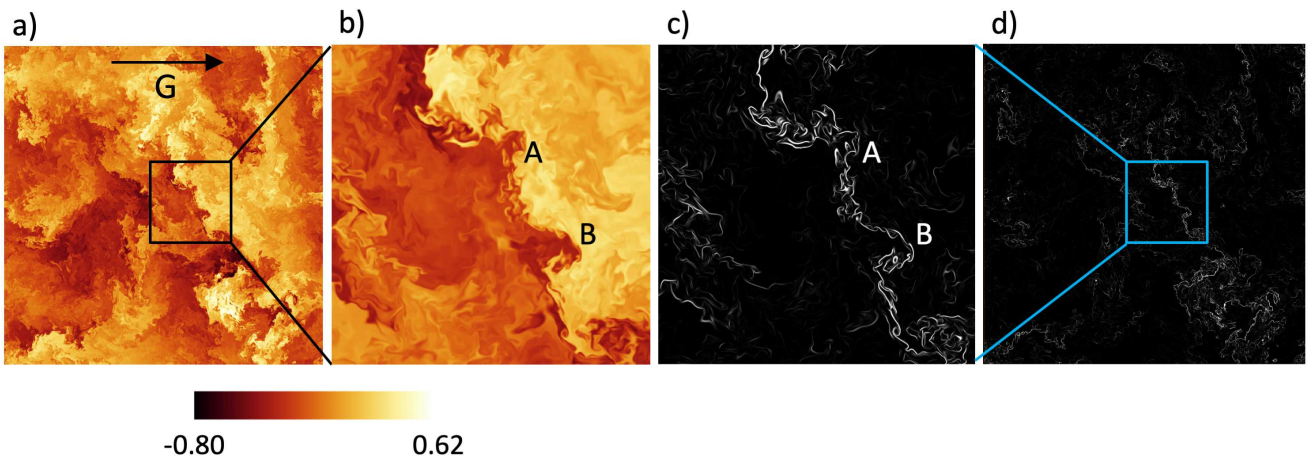}
\caption{A two-dimensional slice of the passive scalar fluctuation field $\theta$ in the homogeneous isotropic turbulent flow. (a) Contours of the cut are shown, as is the direction of the mean scalar gradient $G$; (b) shows the magnification of the contours within the black box in panel (a). A segment of a scalar front is highlighted by A and B. The color bar at the bottom holds for both figures (a) and (b) in units of $GL_0$, which is the maximum available difference of the mean scalar in the box. (c) The zoom of the magnitude of the gradient field, corresponding to (b). Steep gradients appear bright. (d) Full cross section of the scalar gradient, corresponding to the section (a).}
\label{fig1.fig}
\end{figure*}
\begin{figure*}
\includegraphics [width=0.65\textwidth]{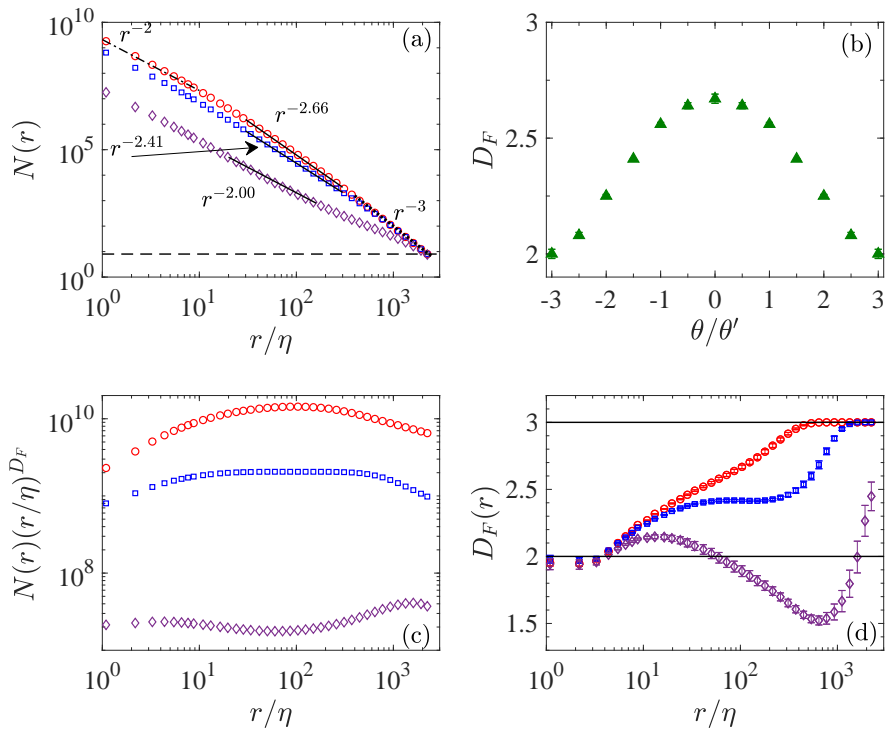}
\caption{Box-counting results for scalar iso-levels for $\tilde\theta = \theta/\theta^\prime$ of 0 (circles), 1.5 (squares) and 3 (diamonds). These symbols have the same meaning in Figs.\ (a), (c) and (d). (a) The number of boxes $N(r)$ required to cover different iso-levels $\thetanr$ versus box size $r$. The box count is evaluated as $N(r) = \la N(r) \ra (L_0/r)^3$, where $\la N(r) \ra$ is the average box number from a gliding-box algorithm \cite{Allain1991}. The box-count obtained by considering non-overlapping boxes is in good agreement with that from the gliding-box algorithm. Dashed horizontal line at $8$ corresponds to $N(r)$ at $r = L_0/2$, which yields the box-counting dimension $D_F = 3$ for $r \sim L_0$, shown by the dotted line. Also shown is the fractal dimension $D_F = 2$ for diffusive scales by the dash-dot line. The power laws in the intermediate scale range are marked by solid lines. (b) The box-counting dimension $D_F$ as a function of the iso-level in the region $|\tilde\theta| \le 3$. (c) Double-logarithmic plot of $N(r)$ compensated by $r^{D_F}$ versus $r/\eta$, where $D_F$ is obtained from (a) using the method of least-squares. (d) Logarithmic local slope $D_F(r) = -d[\log N(r)]/d[\log r]$ of three different iso-levels $\tilde\theta$. Solid horizontal lines drawn for ordinate values $2$ and $3$ denote the small-scale and large-scale dimension limits, respectively.}
\label{nr.fig}
\end{figure*}

%%%%%%%%%%%%%%%%%%%%%%%%%%%%
Figure \ref{nr.fig}(a) shows the box-counting result for three iso-levels, corresponding, respectively, to the mean value of the passive scalar, $\theta=0$, $1.5 \theta^\prime$ away from the mean, and, finally, to $3 \theta^\prime$ away from the mean. For small $r$, the number of boxes $N(r)$ varies as $r^{-2}$, as should be expected for a spatially smooth field. For $r$ close to $L$, $N(r) \sim r^{-3}$, which shows the space-filling character of the scalar front at the largest scales. In an intermediate range of scales of the order of a decade, $N(r) \sim r^{-D_F}$, where $2 \le D_F \le 3$. Only three iso-level sets are shown in Fig~\ref{nr.fig}(a) for reasons of clarity. We will shortly examine the quality of these fits (and the results for iso-level sets), but if we plot the dimension $D_F$ obtained from linear fits in the double-logarithmic plots, against the iso-level values $\tilde\theta \equiv \theta/\theta^\prime$, we find a continuous variation from $2$ for large values of iso-levels to about $2.67$ for the iso-level corresponding to the mean of $\theta$ (see Fig.~\ref{nr.fig}(b)). That the dimension is $D_F\approx 2$ for iso-level sets with large thresholds is obvious because essentially no mixing has taken place that far away from the mean, and hardly any mixing front is available for larger thresholds than about $3\theta^\prime$. We will comment separately on the peak value of the dimension.

%%%%%%%%%%%%%%%%%%%%%%%%%%%%
%%%%%%%%%%%%%%%%%%%%%%%%%%%%

The quality of the power laws has been a matter of contention (see e.g.~discussions in Refs.~\cite{sreenivasan1991, dimotakis2005}), so we explore this issue further, first by showing, in Fig.~\ref{nr.fig} (c), the compensated plots using the $D_F$ values obtained in Fig.~\ref{nr.fig} (a). There is a very clear plateau for $\tilde\theta = 1.5$, for $D_F = 2.35$, as was also found in the past analyses of experimental \cite{sreenivasan1986} and DNS \cite{sangil2001} data; this is also reasonably true for $\tilde\theta = 3$ for which $D_F = 2$ because the scalar with such large deviations from the mean has essentially not mixed, with no chance of developing a contorted front. For $\theta=0$, however, there is at best a hint of a plateau---a point to which we shall return later. In addition, we show in Fig.~\ref{nr.fig} (d) the corresponding local slopes. Again, it is clear that local slopes have a region of satisfactory constancy for $\tilde\theta = 1.5$, perhaps roughly so also for $\tilde\theta = 3$, but possess just a hint of inflection for the iso-level of 0. Incidentally, most past skeptics of power laws have focused on the case $\theta = 0$. The rest of the paper is mostly an effort to understand the results of Fig.~\ref{nr.fig}, and connect them, qualitatively, with the ramp-cliff structure. For determining an iso-level set for a chosen level, we take a small band of scalar values around that level; in the Appendix \label{isolevel.app} we describe how the band thickness was determined.

\section{Upper limit to scaling dimension by geometric measure theory}
We now consider the case of zero iso-level for which, as discussed already, there is only a hint of an inflection in local slope. In Ref.~\cite{constantin1991}, it was shown by geometric measure theory, and the standard hypothesis that velocity increments in classical turbulence are H\"older continuous with an exponent of 1/3, that a scalar interface is indeed a fractal with the dimension of $D_F=2\frac{2}{3}$. By drawing lines in log-log plots as in Fig.~ \ref{nr.fig}, Constantin et al. \cite{constantin1991} deemed that the dimension was supported experimentally \cite{sreenivasan1991} to be $2\frac{2}{3}$. Since, as we have seen, the evidence for it is not as clean as for other iso-levels, we now examine this issue in greater detail. 

\begin{figure}
\includegraphics [width=0.5\textwidth]{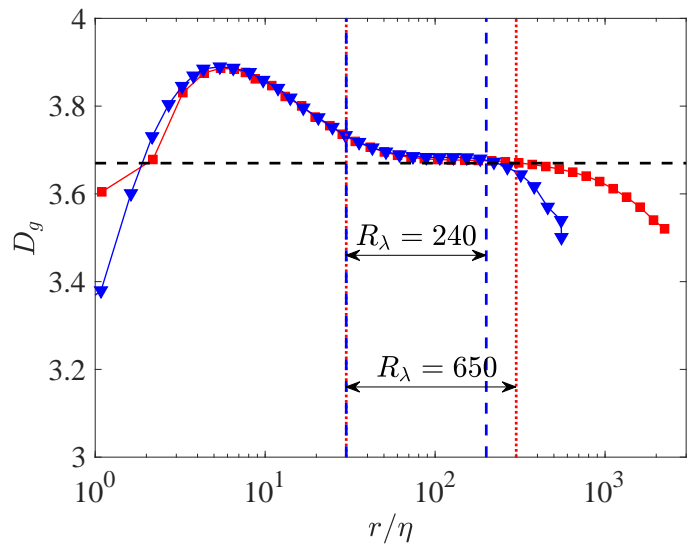}
\caption{Local scaling dimension of the Hausdorff volume $D_g(\tilde r)$ versus $r/\eta$. The quantity is obtained for a passive scalar graph over a three-dimensional ball $B_r$ of varying radius $r$ for $\rel=650$ (squares) and for an additional DNS run at $\rel = 240$ (inverted triangles). In the inertial range (demarcated by the vertical dashed-dot and dashed lines), $D_g = 3.67$, indicated by the dashed line, which is consistent with the fractal dimension of $2.67$ for the zero iso-level shown in Fig.\ \ref{nr.fig}.}
\label{gm.fig}
\end{figure}

We first describe the geometric measure theory \cite{morgan2000} result briefly. The central object of interest is the scaling behavior of the Hausdorff volume $H$ of a passive scalar graph over a three-dimensional ball $B_r$ with a volume $V=4\pi r^3/3$ and a radius $r$ which is given \cite{constantin1991,grossmann1994} by
%---------------------------------------------------------------------------------
\begin{equation}
H(g(B_r)) \sim r^{D_g}\,,
\end{equation}  
%---------------------------------------------------------------------------------
with the graph $g$ over the sphere, defined at a particular time-instant as $g(B_r)=\{({\bm x}, \theta) | {\bm x}\in B_r \;\text{and}\; \theta=\theta({\bm x})\}$. Here, $D_g$ is the scaling dimension of the graph which is by definition connected to the fractal dimension $D_F$ by
%---------------------------------------------------------------------------------
\begin{equation}
D_F=D_g-1\,.
\label{dfdg}
\end{equation}
%---------------------------------------------------------------------------------
For the derivation of $D_g$ we follow Ref. \cite{grossmann1994} (see also \cite{eckhardt1999} for a two-dimensional case). According to the theory, the relative Hausdorff volume is given by
%---------------------------------------------------------------------------------
\begin{equation}
\frac{H(g(B_r))}{V} = \frac{1}{V}\int_{B_r} \sqrt{1+ r^2 |{\bm \nabla} \tilde\theta|^2}\;dV
                                        \le  \sqrt{ 1+ \frac{3}{4\pi r}\int_{B_r} |{\bm \nabla} \tilde\theta|^2 \;dV}\,.
\label{haussd}                                       
\end{equation}  
%---------------------------------------------------------------------------------
The expression in the middle of \eqref{haussd} is a generalization of the calculation formula for the length of a curve. As before, $\tilde\theta=\theta/\theta^\prime$. The second step follows from the Cauchy-Schwarz inequality. As discussed in \cite{constantin1991}, further progress can be made by substituting for the square of the scalar gradient by the terms of the underlying advection-diffusion equation \eqref{scalar} of the passive scalar $\theta$. In the statistically stationary regime, one obtains, by the multiplication of this equation with $\theta$ and a subsequent integration by parts, the following expression:
%---------------------------------------------------------------------------------
\begin{equation}
|{\bm \nabla} \tilde\theta|^2=-\frac{1}{2D} ({\bm u}\cdot{\bm \nabla})\tilde\theta^2  + \frac{1}{2}{\bm \nabla}^2\tilde\theta^2 - \frac{u_x G \tilde\theta}{D\theta^{\prime}}\,.
\label{estimate}
\end{equation}
%---------------------------------------------------------------------------------
In \cite{grossmann1994}, it was shown that the second and third terms on the right hand side of \eqref{estimate} are bounded by the first term. The first term itself can be rewritten as an expression that contains the second-order structure function of longitudinal velocity increment $S_{\parallel}(r)$. In deriving the final result, which is given by
%---------------------------------------------------------------------------------
\begin{equation}
\frac{H(g(B_r))}{V} \le  \sqrt{ 1+ \frac{3 \sqrt{3}}{2} \tilde r \sqrt{\tilde{S}_{\parallel}(\tilde r)}}\,,
\label{final}
\end{equation}  
%---------------------------------------------------------------------------------
one uses the homogeneity of the scalar turbulence, the Cauchy-Schwarz inequality once more, and a scalar flatness of $F_{\tilde{\theta}}=\langle \tilde{\theta}^4\rangle=3$. Here, $\tilde{r} \equiv r/\eta$ and $\tilde{S}_{\parallel} \equiv S_{\parallel}/v_\eta^2$, where $v_\eta = (\nu \langle \epsilon \rangle)^{1/4}$ is the Kolmogorov velocity. Further details on the derivation of the formula can be found in \cite{grossmann1994,eckhardt1999}. From Eq.\ (\ref{final}) follows the local slope 
%---------------------------------------------------------------------------------
\begin{equation}
D_g(\tilde r) = 3 + \frac{\mbox{d}}{\mbox{d} \log \tilde{r}} \log \sqrt{ 1+ \frac{3 \sqrt{3}}{2} \tilde r \sqrt{\tilde{S}_{\parallel}(\tilde r)}}\,,
\label{final1}
\end{equation}  
%---------------------------------------------------------------------------------
where we assume that the inequality can be replaced by an equality. If we assume that the inertial range scaling exponent of the longitudinal structure function to be $\zeta_{\parallel}\approx 2/3$ (as is thought to hold for Kolmogorov turbulence---with slight intermittency correction if needed \cite{Frisch1994}), we find from \eqref{final1} that $D_g=3\frac{2}{3}$. We plot in Fig.~\ref{gm.fig} the results that follow when the structure function from the DNS is inserted. For comparison, we add another data record at $R_{\lambda}=240$. The power-law scaling is not very extensive, but a range of scales certainly exists for which $D_F$ is close to $2\frac{2}{3}$ (indicated by the dashed line at $D_g = 3.67$). Our whole analysis in this section did not make any assumption on the particular level set. We can thus interpret the resulting box-counting dimension given by \eqref{dfdg} as an upper bound $\overline{D}_F$. In other words, a passive scalar in a three-dimensional flow can be stirred and advected in the inertial subrange only to level set with $D_F\le \overline{D}_F$. Our analysis in Fig.~\ref{nr.fig}(b) clearly supports this bound. 

%%%%%%%%%%%%%%%%%%%
%%%%%%%%%%%%%%%%%%%

\section{Unique monofractal scaling in strain-dominated cliff regions}
\begin{figure}
\includegraphics [width=0.5\textwidth]{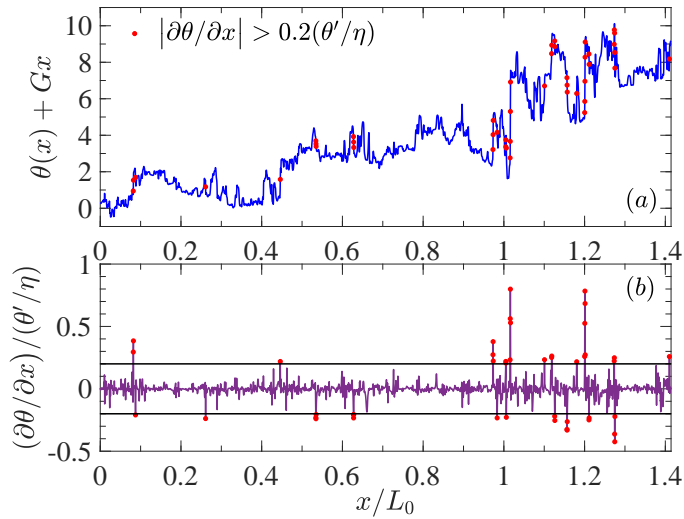}
\caption{Scalar cliff region identification. (a) The line trace taken diagonally across the data shown in Fig.\ \ref{fig1.fig} (a) from the lower left to the upper right corner. Red points indicate the positions at which the derivative magnitude $|\partial \theta/\partial x|$ exceeds the threshold given in the legend of the figure. The criterion is taken from \cite{iyer2018}. The mean scalar gradient is added to show the ramp-cliff structure more clearly. (b) is the corresponding derivative trace $\partial \theta/\partial x$ in units of $\theta^{\prime}/\eta$ along the same diagonal as (a), indicating that the highest amplitudes are captured by this criterion.}
\label{fig4.fig}
\end{figure}

Our box-counting analysis in Sec.~\ref{sec3} revealed that different scalar iso-level sets show different scaling dimensions. We might therefore ask if a unique monofractal can be observed under any circumstances at all. The cliff regions, i.e., the regions in which the magnitude of $\partial \theta/\partial x$ is large, already satisfy this expectation roughly. In Ref. \cite{iyer2018}, we have identified a box-counting dimension of $D_F=1.8$ for the spatial support for this particular subset of the whole volume. Figure \ref{fig4.fig} highlights these regions as red points in a total scalar fluctuation profile $\theta+Gx$ (blue line) taken across the diagonal of Fig.~\ref{fig1.fig}(a). The bottom panel of this figure illustrates the selection criterion by which we identify the scalar derivative with the strongest spatial variations. We found in \cite{iyer2018} that the scalar iso-levels corresponding to these spatial regions have a box-counting dimension of $D_F\le 1.8$, which suggests that the cliffs are loosely in the form of a surface with holes.

\begin{figure}
\includegraphics [width=0.5\textwidth]{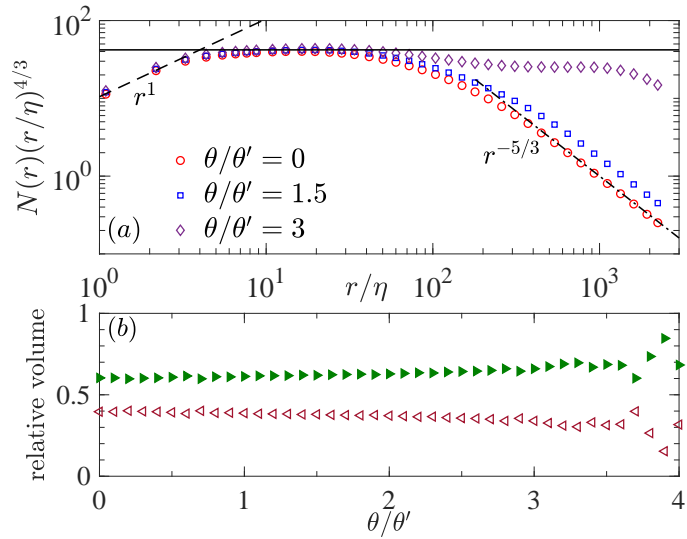}
\caption{Scaling of scalar iso-level conditioned to high strain on the cliffs. (a) The compensated double-logarithmic plot of $N(r)$ versus size $r$, for the strain-dominated regions of the flow. The plateau region marked by the solid line corresponds to the $4/3$-scaling and is exhibited by all the iso-levels sets shown. The small and large scale regimes exhibit a $-1/3$ (dashed line) and $-3$ (dash-dot line) scaling dependency, respectively. (b) Relative volume fraction of strain-dominated (filled triangles) and rotation-dominated regions (open triangles) for the chosen scalar iso-level set $\thetanr$. This analysis relates to the strain-dominated cliff regions only with ($\scgx > 0.2 \scglarge$).}
\label{fig5.fig}
\end{figure}

But one can do better in terms of the quality of scaling by restricting attention on cliff regions connected to a persistent local straining motion, a known process studied in the chaotic mixing regime of high-Schmidt-number turbulence \cite{Kushnir2006,Villermaux2019,Goetzfried2019}. For this purpose, we refine the analysis and examine strain-dominated subsets in the cliff regions. They are extracted from a local eigenvalue analysis of the velocity gradient tensor ${\bm \nabla}{\bm u}$ (grid point by grid point); see \cite{Ashurst1987,JS05}. The dominance of local pure strain (as opposed to local rotation) implies that the velocity gradient tensor is locally symmetric and possesses three real eigenvalues that sum up to zero due to incompressibility. Box-counting results for these regions are shown in Fig.~\ref{fig5.fig}. We find in panel (a) that for all iso-levels the scaling is uniformly the same and approximately $4/3$, suggesting that the strain dominated regions of the cliff are better regarded as highly convoluted line-like objects rather than surfaces full of holes. Figure \ref{fig5.fig} (b) shows the relative volumes of the strain- and rotation-dominated regions in the spatial support of the cliffs which have to sum up to unity; these results are the outcome of the eigenvalue analysis of the velocity gradient tensor. It is seen that the volume fractions do not change much with the value of the iso-level. The conclusion is that in strain-dominated regions of the spatial support of the cliffs, there is a unique fractal scaling dimension for all iso-level sets, and its value is approximately 4/3. Such a box-counting dimension could correspond to material lines that are most probably stirred by velocity increments in the inertial range, characterized by the spatial scaling of $r^{1/3}$. 

%%%%%%%% END %%%%%%%%%%%%%%
%%%%%%%%%%%%%%%%%%%%%%%%%%%%%

Finally, we may now turn to the physical meaning of the upper bound of about 2.67 for the fractal dimension of the iso-scalar surfaces, which corresponds to $\theta = 0$. This shows that such levels sets are not space filling in the inertial range. If perfect mixing occurs at the inertial-range scales, such a surface would have a space-filling dimension of 3. Given that the regions where mixing has been accomplished on inertial-range scales is only $2\frac{2}{3}$, we conclude that there is an upper bound to the mixing in turbulent flows \cite{sreeni2019}. It would be rewarding to prove this result analytically. The observation of strong ramp-cliff structures at even the highest Reynolds numbers considered is completely consistent with this view of incomplete mixing with a finite bound.

\section{Conclusions}
We have conducted a geometric analysis of passive scalar iso-level sets in three-dimensional turbulence at high Reynolds numbers and a Schmidt number ${\rm Sc}=1$. The homogeneous and isotropic box turbulence advecting the flow is characterized by an inertial range over an order of magnitude in which the Kolmogorov 4/5-ths law holds, as shown in Ref.\ \cite{iyer2018}. Furthermore, the Kolmogorov scale $\eta$ is resolved with one grid spacing, which provides a high-quality DNS data set as the basis of analysis. 

We have shown that a box-counting scaling dimension $D_F$ can be obtained for all iso-levels, excepting those for high amplitudes, say $|\tilde\theta| > 3$, because there is essentially no mixing at such high iso-levels and the front, such as may exist, has very little likelihood of developing any contortions that lead to fractal scaling. The box-counting dimension $D_F$ varies with the iso-level magnitude.

By means of geometric measure theory, we derived an upper bound $D_F\le 8/3$ which is the maximum possible dimension of the iso-level sets; this corresponds to the iso-level set of zero, towards which all mixing processes are driven. If the mixing were complete, the zero iso-level sets would be space-filling and we would obtain a dimension of 3. The fact that we do not achieve this condition suggests that the mixing is not complete in a turbulent flow. This is because there is a finite probability of encountering cliffs across which the scalar jumps by almost the amount allowed in the flow. Expressed differently, there are always positions in the flow where the lowest concentrations of the scalar are separated by the highest concentration levels only by the smallest scale available to the flow. That this happens for the case of homogeneous and isotropic turbulence suggests that it must be a general feature of turbulence, which leads us to conclude that there is an upper bound to turbulent mixing in practice. 

We already noted that the box-counting dimension $D_F$ varies with the iso-level magnitude and that a unique monofractal behavior with a scaling dimension independent of the iso-level is not obtainable. However, such a unique monofractal scaling of scalar iso-levels can be obtained when two additional conditions are imposed: (1) select those points of space that spatially support the steep scalar cliffs, and (2) condition the box-counting analysis of iso-levels on this support to high-strain events. In some sense, this is the backbone of structures that prevent complete scalar mixing.

An extension of this analysis for high-Schmidt-number passive scalar turbulence can be considered as the natural next step. This study is currently under way and will be reported elsewhere.  
 
\acknowledgments
The computations and data analyses reported in this paper were performed using advanced computational facilities provided by the Texas Advanced Computation Center (TACC) under the XSEDE program supported by NSF.  The datasets used were originally generated using supercomputing resources at the Oak Ridge Leadership Computing Facility at the US Department of Energy Oak Ridge National Laboratory. JS wishes to thank the Tandon School of Engineering at New York University for financial support. KRS thanks Dr. Inigo San Gil who made the same fractal analysis in 2001 on a $512^3$ data set.

\appendix
\section{The definition of the scalar iso-level thickness}
%%%%%%%%%%%%%%%%%%%%5
\def\del{\Delta}
\def\diff{D}
\def\sch{\textrm{Sc}}
\def\epsd{\epsilon_\theta}
\def\dth{\delta \theta}
\def\kresc{{k_{max}\eta_B}}
\def\kmax{k_{\textrm{max}}}
\def\bsc{\eta_B}
\def\const{\frac{2\sqrt{2}\pi}{3C}}
\def\trms{\theta^\prime}
\def\dc{A(1+\sqrt{1+(B/\rel)^2})}
\def\dcsq{\Big[A(1+\sqrt{1+(B/\rel)^2})\Big]^{1/2}}
\label{isolevel.app}
\noindent
Consider passive scalar fluctuation $\theta$ with diffusivity $\diff$, mean scalar dissipation $\epsd$ and Schmidt number $\sch = \nu/\diff$ where $\nu$ is the kinematic viscosity of the advecting fluid. The typical scalar variation across grid cell $\del = L_0/N$ in a cube with edge length $L_0$ with $N$ points to a side is
\beq
\label{dthetares.eq}
\dth = \Big( \frac{\epsd}{\diff} \Big)^{1/2} \del \;.
\eeq 
Denoting the small-scale resolution parameter $\kresc$ by $C$, where $\kmax$ is the highest resolvable wavenumber in a $N^3$ simulation with smallest non-zero wavenumber magnitude $k_0 = 2\pi/L_0$ and $\eta_B$ is the Batchelor scale $\bsc = \eta/\textrm{Sc}^{1/2}$, we can write
\beq
\label{kres.eq}
C = \frac{\sqrt{2}}{3} N k_0 \bsc = \frac{\sqrt{2}}{3} \Big(\frac{L_0}{\del}\Big) \Big(\frac{2\pi}{L_0} \Big ) \bsc. 
\eeq
Solving for $\del$ in above equation, substituting into Eq.~\ref{dthetares.eq} and dividing both sides by the root-mean-square scalar fluctuation $\trms$ we get 
\beq
\label{dthetanew.eq}
\frac{\dth}{\trms} = \Big( \frac{\epsd}{\diff} \Big)^{1/2} \frac{1}{\trms} \const \bsc \;.
\eeq
Assuming dissipative anomaly for the scalar field in isotropic turbulence \cite{DSY05} we can write
\beq
\label{chid.eq}
\epsd = \dc \frac{{\trms}^2 u^\prime}{L} \;,
\eeq
where $u^\prime$ is the root-mean-square velocity fluctuation, $L$ is the flow integral scale and $A$ and $B$ are constants that depend on the Schmidt number \cite{DSY05}.
%$A^\prime \app 0.4$ and $B^\prime \app 31$ for $\sch = 1$  \cite{DSY05}.
Substituting Eq.~\ref{chid.eq} into Eq.~\ref{dthetanew.eq} and rearranging we get 
\beq
\label{dthetanext.eq}
\frac{\dth}{\trms} = \const \dcsq \Big( Re \; \sch \Big)^{1/2} \Big (\frac{\eta_B}{L} \Big ) \;,
\eeq
where $Re$ and $\rel$ denote the Reynolds numbers based on the integral scale and the Taylor microscale, respectively, and are related to each other in isotropic turbulence as
$\rel = (\frac{20}{3} Re)^{1/2}$. For $Sc = 1$, $\eta_B = \eta$ and thus we can finally write 
\beq
\label{dfinal.eq}
\frac{\dth}{\trms} = \const \dcsq \Big (\frac{20}{3} \Big )^{1/4} \rel^{-1/2} \;.
\eeq
For a resolution of $N^3 = 4096^3$ in our DNS with $\rel = 650$ and $C = 2.72$, substituting $A \app 0.4$ and $B \app 31$, it follows that the iso-level thickness is effectively
\beq
\label{dnsval.eq}
\pm\dth/\trms \app \pm 0.03\,,
\eeq 
for the present data.

\end{document}